\PassOptionsToPackage{unicode}{hyperref}
\PassOptionsToPackage{hyphens}{url}
\PassOptionsToPackage{dvipsnames,svgnames,x11names}{xcolor}
\documentclass[
]{article}
\usepackage{amsmath,amssymb}
\usepackage{iftex}
\ifPDFTeX
  \usepackage[T1]{fontenc}
  \usepackage[utf8]{inputenc}
  \usepackage{textcomp} % provide euro and other symbols
\else % if luatex or xetex
  \usepackage{unicode-math} % this also loads fontspec
  \defaultfontfeatures{Scale=MatchLowercase}
  \defaultfontfeatures[\rmfamily]{Ligatures=TeX,Scale=1}
\fi
\usepackage{lmodern}
\ifPDFTeX\else
\fi
\IfFileExists{upquote.sty}{\usepackage{upquote}}{}
\IfFileExists{microtype.sty}{% use microtype if available
  \usepackage[]{microtype}
  \UseMicrotypeSet[protrusion]{basicmath} % disable protrusion for tt fonts
}{}
\makeatletter
\@ifundefined{KOMAClassName}{% if non-KOMA class
  \IfFileExists{parskip.sty}{%
    \usepackage{parskip}
  }{% else
    \setlength{\parindent}{0pt}
    \setlength{\parskip}{6pt plus 2pt minus 1pt}}
}{% if KOMA class
  \KOMAoptions{parskip=half}}
\makeatother
\usepackage{xcolor}
\usepackage[margin=1in]{geometry}
\usepackage{color}
\usepackage{fancyvrb}

\DefineVerbatimEnvironment{Highlighting}{Verbatim}{commandchars=\\\{\}}
\usepackage{framed}
\definecolor{shadecolor}{RGB}{248,248,248}
\newenvironment{Shaded}{\begin{snugshade}}{\end{snugshade}}

\newcommand{\AttributeTok}[1]{\textcolor[rgb]{0.13,0.29,0.53}{#1}}

\newcommand{\CommentTok}[1]{\textcolor[rgb]{0.56,0.35,0.01}{\textit{#1}}}

\newcommand{\ConstantTok}[1]{\textcolor[rgb]{0.56,0.35,0.01}{#1}}
\newcommand{\ControlFlowTok}[1]{\textcolor[rgb]{0.13,0.29,0.53}{\textbf{#1}}}

\newcommand{\DecValTok}[1]{\textcolor[rgb]{0.00,0.00,0.81}{#1}}

\newcommand{\FloatTok}[1]{\textcolor[rgb]{0.00,0.00,0.81}{#1}}
\newcommand{\FunctionTok}[1]{\textcolor[rgb]{0.13,0.29,0.53}{\textbf{#1}}}

\newcommand{\NormalTok}[1]{#1}

\newcommand{\OtherTok}[1]{\textcolor[rgb]{0.56,0.35,0.01}{#1}}

\newcommand{\SpecialCharTok}[1]{\textcolor[rgb]{0.81,0.36,0.00}{\textbf{#1}}}

\newcommand{\StringTok}[1]{\textcolor[rgb]{0.31,0.60,0.02}{#1}}

\usepackage{longtable,booktabs,array}
\usepackage{calc} % for calculating minipage widths
\usepackage{etoolbox}
\makeatletter
\patchcmd\longtable{\par}{\if@noskipsec\mbox{}\fi\par}{}{}
\makeatother
\IfFileExists{footnotehyper.sty}{\usepackage{footnotehyper}}{\usepackage{footnote}}
\makesavenoteenv{longtable}
\usepackage{graphicx}
\makeatletter
\def\maxwidth{\ifdim\Gin@nat@width>\linewidth\linewidth\else\Gin@nat@width\fi}
\def\maxheight{\ifdim\Gin@nat@height>\textheight\textheight\else\Gin@nat@height\fi}
\makeatother
\setkeys{Gin}{width=\maxwidth,height=\maxheight,keepaspectratio}
\makeatletter
\def\fps@figure{htbp}
\makeatother
\providecommand{\tightlist}{%
  \setlength{\itemsep}{0pt}\setlength{\parskip}{0pt}}
\ifLuaTeX
\usepackage[bidi=basic]{babel}
\else
\usepackage[bidi=default]{babel}
\fi
\babelprovide[main,import]{american}
\def\languageshorthands#1{}
\usepackage{hyperref}
\definecolor{linkcolor}{HTML}{D55E00}
\definecolor{citecolor}{HTML}{009E73}
\definecolor{urlcolor}{HTML}{0072B2}
\hypersetup{
    colorlinks,
    linkcolor={linkcolor},
    citecolor={citecolor},
    urlcolor={urlcolor}
}
\usepackage{xcolor}
\definecolor{good}{HTML}{009E73}
\definecolor{bad}{HTML}{D55E00}

\makeatletter
\@ifundefined{DecValTok}{}{
  \renewcommand{\DecValTok}[1]{\textcolor[HTML]{009E73}{#1}}
  \renewcommand{\FloatTok}[1]{\textcolor[HTML]{009E73}{#1}}
  \renewcommand{\ConstantTok}[1]{\textcolor[HTML]{009E73}{#1}}
  \renewcommand{\ControlFlowTok}[1]{\textcolor[HTML]{0072B2}{\textbf{#1}}}
  \renewcommand{\OtherTok}[1]{\textcolor[HTML]{000000}{#1}}
  \renewcommand{\CommentTok}[1]{\textcolor[HTML]{999999}{\textit{#1}}}
  \renewcommand{\AttributeTok}[1]{\textcolor[HTML]{CC79A7}{#1}}
  \renewcommand{\FunctionTok}[1]{\textcolor[HTML]{56B4E9}{#1}}
}
\makeatother

\usepackage{caption}
\usepackage{enumitem}
\setlist[itemize]{topsep=-5pt}
\setlist[enumerate]{topsep=-5pt}

\usepackage[absolute]{textpos}

\usepackage[framemethod=TikZ]{mdframed}
\definecolor{examplecolor}{HTML}{999999}
\surroundwithmdframed[
  leftmargin=\parindent,
  skipabove=\medskipamount,
  skipbelow=\medskipamount,
  linecolor=examplecolor,
  linewidth=1pt,
  roundcorner=5pt
]{example}
\BeforeBeginEnvironment{document}{
  \counterwithout{example}{section}
}
\newcommand{\x}{\mathbf{x}}
\newcommand{\N}{\mathcal{N}}

\newcommand{\KDE}{\operatorname{KDE}}
\newcommand{\HF}{\operatorname{HF}}
\newcommand{\HFS}{\operatorname{HF7}}
\newcommand{\hQ}{\hat{Q}}
\newcommand{\hf}{\hat{f}}
\newcommand{\hF}{\hat{F}}
\newcommand{\HD}{\operatorname{HD}}
\newcommand{\QHF}{\hQ_{\HF}}
\newcommand{\QHFS}{\hQ_{\HFS}}
\newcommand{\QHD}{\hQ_{\HD}}
\newcommand{\QTHD}{\hQ_{\operatorname{THD}}}
\newcommand{\fHD}{\hf_{\HD}}
\usepackage{amsmath}
\usepackage{float}
\usepackage{csquotes}
\usepackage{booktabs}
\usepackage{longtable}
\usepackage{array}
\usepackage{multirow}
\usepackage{wrapfig}
\usepackage{colortbl}
\usepackage{pdflscape}
\usepackage{tabu}
\usepackage{threeparttable}
\usepackage{threeparttablex}
\usepackage[normalem]{ulem}
\usepackage{makecell}
\usepackage{xcolor}
\ifLuaTeX
  \usepackage{selnolig}  % disable illegal ligatures
\fi
\usepackage[style=alphabetic,sorting=anyt]{biblatex}
\usepackage{bookmark}
\IfFileExists{xurl.sty}{\usepackage{xurl}}{} % add URL line breaks if available
\hypersetup{
  pdfauthor={Andrey Akinshin},
  pdflang={en-US},
  colorlinks=true,
  linkcolor={linkcolor},
  filecolor={Maroon},
  citecolor={citecolor},
  urlcolor={urlcolor},
  pdfcreator={LaTeX via pandoc}}

\title{Quantile-respectful density estimation based on

the Harrell--Davis quantile estimator}
\author{Andrey Akinshin\\
JetBrains, \href{mailto:andrey.akinshin@gmail.com}{\nolinkurl{andrey.akinshin@gmail.com}}}
\date{}

\usepackage{amsthm}

\theoremstyle{definition}

\theoremstyle{definition}
\newtheorem{example}{Example}[section]
\theoremstyle{definition}

\theoremstyle{definition}

\theoremstyle{remark}

\begin{document}
\maketitle
\begin{abstract}
Traditional density and quantile estimators are often inconsistent with each other.
Their simultaneous usage may lead to inconsistent results.
To address this issue, we propose a novel smooth density estimator
that is naturally consistent with the Harrell--Davis quantile estimator.
We also provide a jittering implementation to support discrete-continuous mixture distributions.

\end{abstract}

\section{Introduction}\label{sec:intro}

\begin{figure}[H]

{\centering \includegraphics{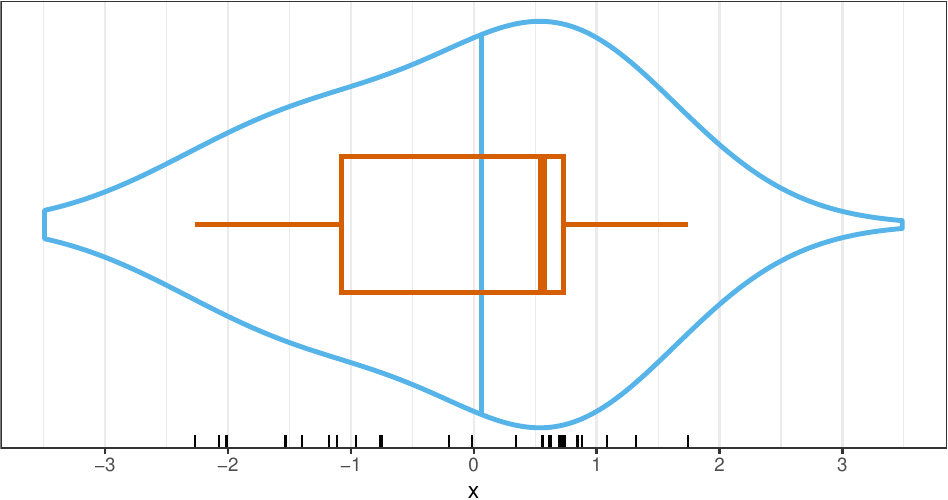} 

}

\caption{A violin plot with the KDE median.}\label{fig:intro}
\end{figure}

A continuous distribution can be described by its probability density function (PDF)~\(f\) or quantile function~\(Q\).
These representations are mutually defined: \(f=\left(Q^{-1}\right)'\).
We consider the problem of building estimations \(\hf\) and \(\hQ\)
based on the given sample \(\x = ( x_1, x_2, \ldots, x_n )\).
In general, the consistency of \(\hf\) and \(\hQ\) is not guaranteed and depends on the chosen estimators.
If these estimators are diverged, the simultaneous usage of \(\hf\) and \(\hQ\) may be error-prone and misleading.
Let us illustrate this problem with a simple example.
In Figure~\ref{fig:intro},
we can see a violin plot of a sample of 30 elements from the standard normal distribution \(\N(0, 1)\).

\clearpage

The density plot is based on the kernel density estimator (KDE, see \autocite{rosenblatt1956})
with the normal kernel and the bandwidth of~0.9.
The boxplot quantiles are estimated by the traditional quantile estimator
(Type~7 in the Hyndman--Fan taxonomy, see \autocite{hyndman1996}).
Both approaches are standard and provided by default in most statistical packages.
Let us denote these estimators by \(\hf_{\KDE}\) and \(\QHFS\).
We also denote the quantile function of a distribution defined by \(\hf_{\KDE}\) as \(\hQ_{\KDE}\).
The median of \(\QHFS\) is presented as a vertical line inside the boxplot.
The classic density plot is enhanced with an additional vertical line representing the median of \(\hat{Q}_{\KDE}\).
The default view is extended by a rug plot on the bottom to show the raw sample values.

As we can see, the median estimations diverge: \(\QHFS(\x) \approx 0.56\), \(\hQ_{\KDE}(\x) \approx 0.06\).
The difference between the medians is about~\(0.5\), which is half of the standard deviation.
Usually, the KDE median is not presented in the plot,
but an experienced researcher can make a good guess about its location.
The violin plot explicitly highlights the sample median, which is inconsistent with the KDE shape.
The density and boxplot parts of the violin plot contradict each other while they are supposed to complement each other.

This inconsistency within the violin plot exemplifies a broader issue encountered in statistical analyses.
If some parts of the research are based on sample quantiles and other parts are based on density estimations,
the results may be misleading.
If we cannot avoid the need for simultaneous usage of \(\hf\) and \(\hQ\), we need to find a way to make them consistent.

One option that could be considered is evaluating quantiles based on KDE.
This approach may lead to inconvenient quantile values.
The default kernel is often the normal one, which is defined on \([-\infty;\infty]\).
This leads to a significant distortion of the extreme quantiles on small samples.
For example, an estimation of the~\(99^\textrm{th}\) percentile can significantly exceed the maximum value of the sample.
Meanwhile, a typical requirement for \(\hQ\) is to provide quantile estimations within the range of the sample values.
In this case, a KDE-based quantile estimator is not an appropriate option.

Another way is to derive the density estimation from the sample quantiles.
Let us call this approach \emph{Quantile-Respectful Density Estimation} or QRDE.
Traditional quantile estimators (which we denote by \(\QHF\)) are based
on a single order statistic or a linear combination of two subsequent order statistics.
Therefore, \(\QHF\) is a piecewise linear function.
The cumulative distribution function (CDF)~\(F\) is an inversion of \(Q\).
Thus, \(\hF\) is also a piecewise linear function.
Since the consistency requirement implies that \(\hf = \hF'\), the density estimation based on \(\QHF\) is a step function
(we call it QRDE-HF).
Another typical requirement of \(\hf\) is smoothness.
The step function does not satisfy this requirement.

To work around the smoothness problem, we can consider alternative quantile estimators.
The focus of this paper is on the Harrell--Davis quantile estimator (see \autocite{harrell1982}), which we denote by \(\QHD\).
Like \(\QHF\), \(\QHD\) always provides estimations in the range of the sample values.
\(\QHD\) is often more efficient than \(\QHF\) (especially for the middle quantiles under light-tailed distributions).
\(\QHD\) is smooth, which allows us to build a smooth and consistent density estimator \(\fHD\), which we call QRDE-HD.
The estimated \(\fHD\) function does not require a bandwidth value: it naturally adapts to the sample size.

In this paper, we develop the concept of \(\fHD\) and show how the mentioned properties of \(\QHD\)
make the proposed approach an advantageous alternative to the classic methods in practical applications.
While \(\fHD\) can be a handy tool in some cases, it does not fit all possible problems.
We also discuss the disadvantages and limitations of \(\fHD\) to make its applicability area clear.

The paper is organized as follows.
In Section~\ref{sec:qrde}, we introduce a new density estimator and explore its properties.
In Section~\ref{sec:jit}, we explain how to properly handle tied values.
In Section~\ref{sec:summary}, we summarize all the results.
In Appendix~\ref{sec:refimpl}, we provide a reference R implementation of the suggested density estimator.

\clearpage

\section{Quantile-respectful density estimation}\label{sec:qrde}

Let us recall the definition of the Harrell-Davis quantile estimator (\autocite{harrell1982}).
The estimation of the \(p^\textrm{th}\) quantile of a sample \(\x\) is defined as follows:

\[
\QHD(\x, p) = \sum_{i=1}^{n} W_{\HD,i} \cdot x_{(i)},\quad
W_{\HD,i} = I_{i/n}(\alpha, \beta) - I_{(i-1)/n}(\alpha, \beta),
\]

where
\(x_{(i)}\) is the \(i^\textrm{th}\) order statistics of \(\x\),
\(I_t(\alpha, \beta)\) is the regularized incomplete beta function,
\(\alpha = (n+1)p\), \(\;\beta = (n+1)(1-p)\).

We want to build a density estimator \(\fHD\),
which defines a density function consistent with the estimated \(\QHD\) quantiles:

\[
\fHD = \left(\QHD^{-1}\right)'.
\]

Accurately estimating the exact value of \(\fHD(x)\) for a specific \(x\) can be challenging and often ineffective.
Therefore, for practical application, we suggest using a pseudo-histogram approximation.
Let us split the \([0;1]\) range into \(k\) intervals of equal size \(1/k=\xi\)
by the cut points denoted by \(p_0, p_1, \ldots p_k\), where \(p_i = i\xi\).
Thus, the \(i^\textrm{th}\) interval is \([p_{i-1}; p_i] = [(i-1)\xi; i\xi]\).
For each of these intervals, we draw a histogram bin.
The left and right boundaries of the \(i^\textrm{th}\) bin are given by \(\QHD(p_{i-1})\) and \(\QHD(p_i)\).
The true area under the corresponding PDF part can be easily calculated:

\[
\int_{Q(p_{i-1})}^{Q(p_i)} f(x) \, dx =
  F(Q(p_i))-F(Q(p_{i-1})) =
  p_i - p_{i-1} =
  i\xi - (i-1)\xi =
  \xi.
\]

Since we have the bin width of \(\QHD(p_i)-\QHD(p_{i-1})\) and the bin area of \(\xi\),
we can derive the height \(h_i\) of the \(i^\textrm{th}\) bin:

\[
h_i = \frac{\xi}{\QHD(p_i)-\QHD(p_{i-1})}.
\]

In practical applications, we recommend using \(\xi=0.001\) in order to obtain a reasonably accurate approximation.
If better computational performance is needed, \(\xi=0.01\) gives a recognizable density shape,
but a step function pattern is noticeable.

The described scheme can be applied to build a density estimation based on any quantile estimators.
For example, we can use traditional \(\QHF\) estimators or some other advanced estimators like
the Sfakianakis-Verginis (\autocite{sfakianakis2008}) and the Navruz-Özdemir (\autocite{navruz2020}) quantile estimators.
However, we advocate using \(\QHD\): thanks to its smoothness and efficiency,
it gives a sophisticated density estimation in most cases.

We can also consider a weighted modification of the Harrell-Davis quantile estimator,
which allows estimating quantiles for weighted samples (see \autocite{akinshin2023wqe}).
Therefore, we can naturally obtain the corresponding weighted QRDE-HD without any modifications in the above approach.

It's important to understand that there is no perfect density estimator.
Each estimator has its advantages and disadvantages and its area of applicability.
QRDE-HD may be useful in some situations, but it does not suit all the problems.
To develop intuition on how to choose a proper estimator, we are going to review a series of examples.
Each example compares QRDE-HD with another density estimator on different data samples.
We will highlight the important properties of QRDE-HD that can help to make a meaningful choice
about the most appropriate estimator in the given context.

\clearpage

\begin{example}[QRDE-HD vs. QRDE-HF7]
\protect\hypertarget{exm:hf7}{}\label{exm:hf7}

In Figure~\ref{fig:hf7}, we can see a comparison of QRDE-HD and QRDE-HF7 on samples of size 10 and 13 respectively
from the standard normal distribution.

The traditional quantile estimators like \(\QHF\) are piecewise linear functions.
Therefore, we can't estimate the density function values at the points where \(\QHF\) is not differentiable.
If we try to build this function numerically according to the above scheme,
we will observe huge spikes at these points.
Thus, the QRDE approach based on the traditional quantile estimators is not practically useful.

In contrast, the Harrell-Davis quantile estimator is a smooth function.
Therefore, we can build a smooth density estimation based on it.
From the first look, QRDE-HD looks good on the presented small samples:
it is smooth,
it has a high density around data points and a low density in the gaps,
and it does not exceed the sample range.

\begin{figure}[H]

{\centering \includegraphics{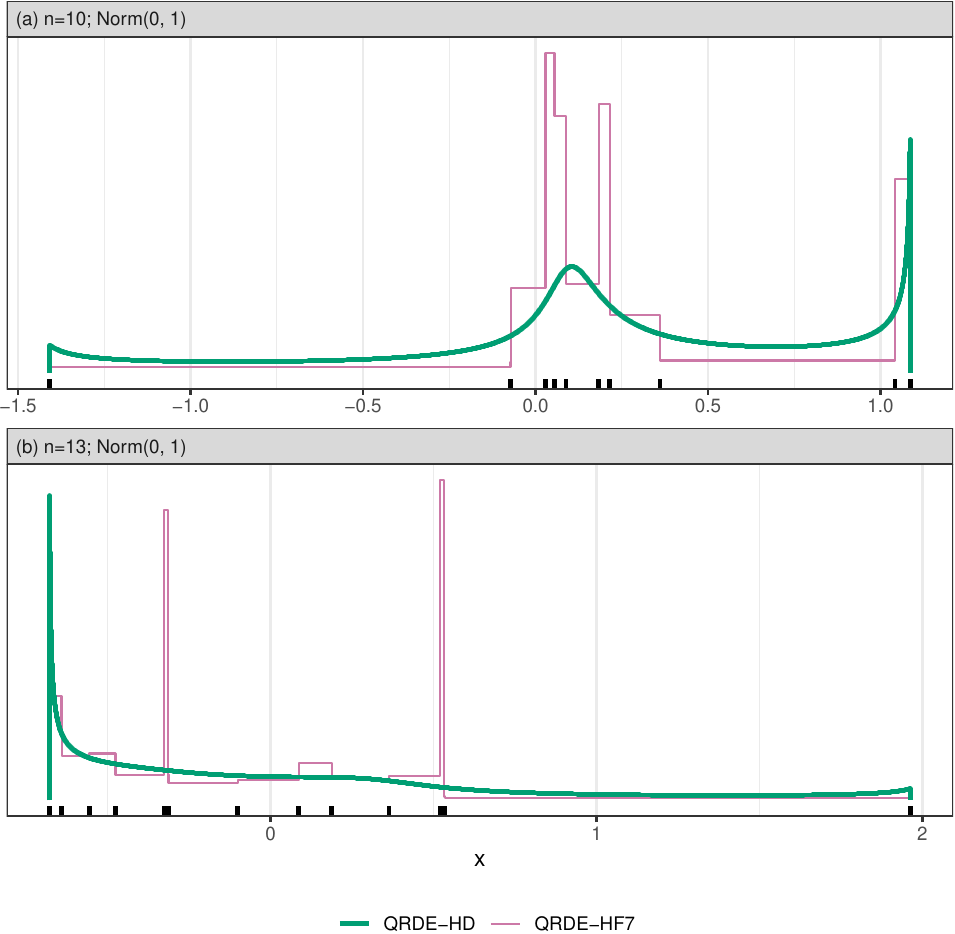} 

}

\caption{QRDE-HD vs. QRDE-HF7.}\label{fig:hf7}
\end{figure}

\end{example}

\clearpage

\begin{example}[QRDE-HD vs. Equal-width Histogram]
\protect\hypertarget{exm:hist}{}\label{exm:hist}

Equal-width histogram is a classic density estimator.
Thanks to its simplicity, it is widely used in practice.
It accurately describes the sample without applying any approximation methods.
However, it has some disadvantages.

The first problem is a lack of smoothness.
In Figure~\ref{fig:hist} (a),
we can see a comparison of QRDE-HD and equal-width histogram on a random sample of medium size.
Both QRDE-HD and the histogram similarly acknowledge the shape of the data,
but QRDE-HD provides a nice-looking smooth curve, while the histogram is a step function.

Another problem is about choosing the proper bin width.
Wide bin widths lead to over-smoothing, while narrow bin widths lead to overfitting.
In Figure~\ref{fig:hist} (b), we have a sample from a distribution with three modes.
The gap between the first and the second modes is much larger than the gap between the second and the third modes.
With the equal-width histogram, the bin width will be defined by the dominant first gap.
As we can see, it turned out to be too wide and, therefore, the last two modes are covered by the same bin.
Meanwhile, QRDE-HD accurately adapts to each local area and properly distinguishes the modes
(assuming the sample size is large enough).

\begin{figure}[H]

{\centering \includegraphics{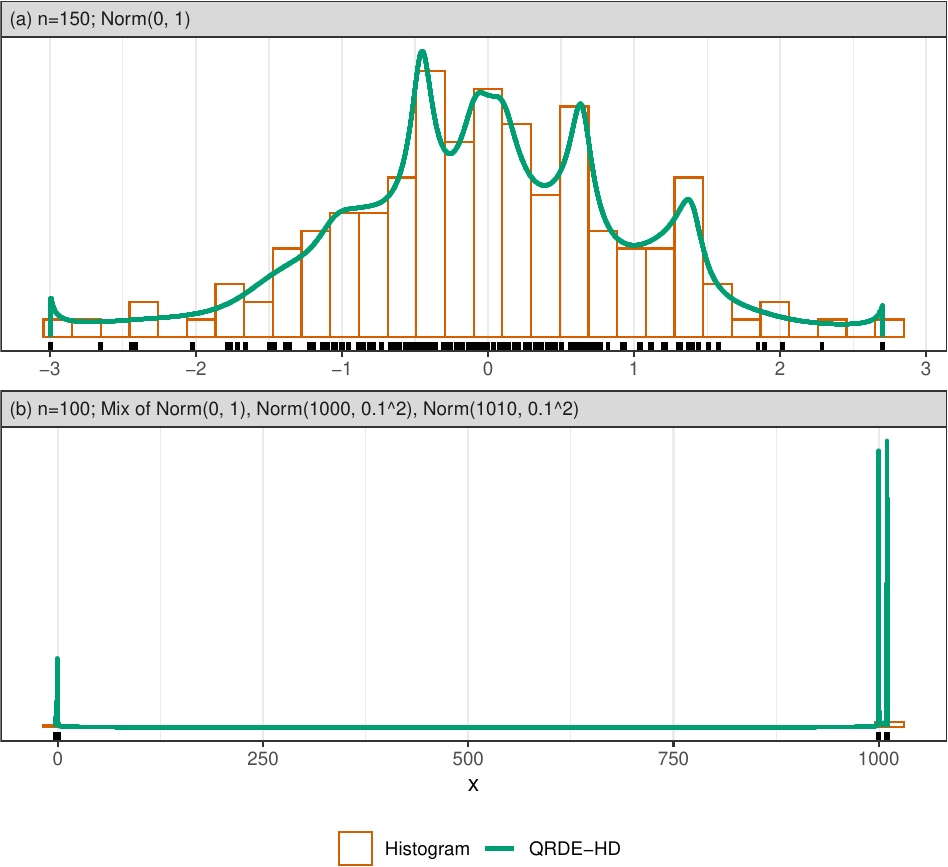} 

}

\caption{QRDE-HD vs. Equal-width Histogram.}\label{fig:hist}
\end{figure}

\end{example}

\clearpage

\begin{example}[QRDE-HD and boundary spikes]
\protect\hypertarget{exm:tail}{}\label{exm:tail}

In Figure~\ref{fig:tail}, we present QRDE-HD (and KDE as a reference) for two uniform cases.
The plots distinctly show a special artifact of QRDE-HD: pronounced spikes at the sample boundaries.
While such spikes could be perceived as an incorrect tail estimation, they reveal the true nature of \(\QHD\).
The intuition behind his phenomena is the following: at the right neighborhood of \(\min(\x)\),
\(x_{(1)}\) has a much stronger impact on the estimation than all the other elements together
due to the shape of the Beta distribution.

While the Harrell-Davis quantile estimator provides accurate estimations of the middle quantiles,
it performs poorly in extreme cases.
The observed spikes should be treated as a peculiar artifact of \(\QHD\) rather than a phenomenon in the data.
With enough experience working with \(\QHD\), such spikes become convenient and do not cause confusion.
However, when a QRDE-HD is presented to people without a relevant background,
it may make sense to present only the middle part (e.g.~\(0.1 < p < 0.9\)).

\begin{figure}[H]

{\centering \includegraphics{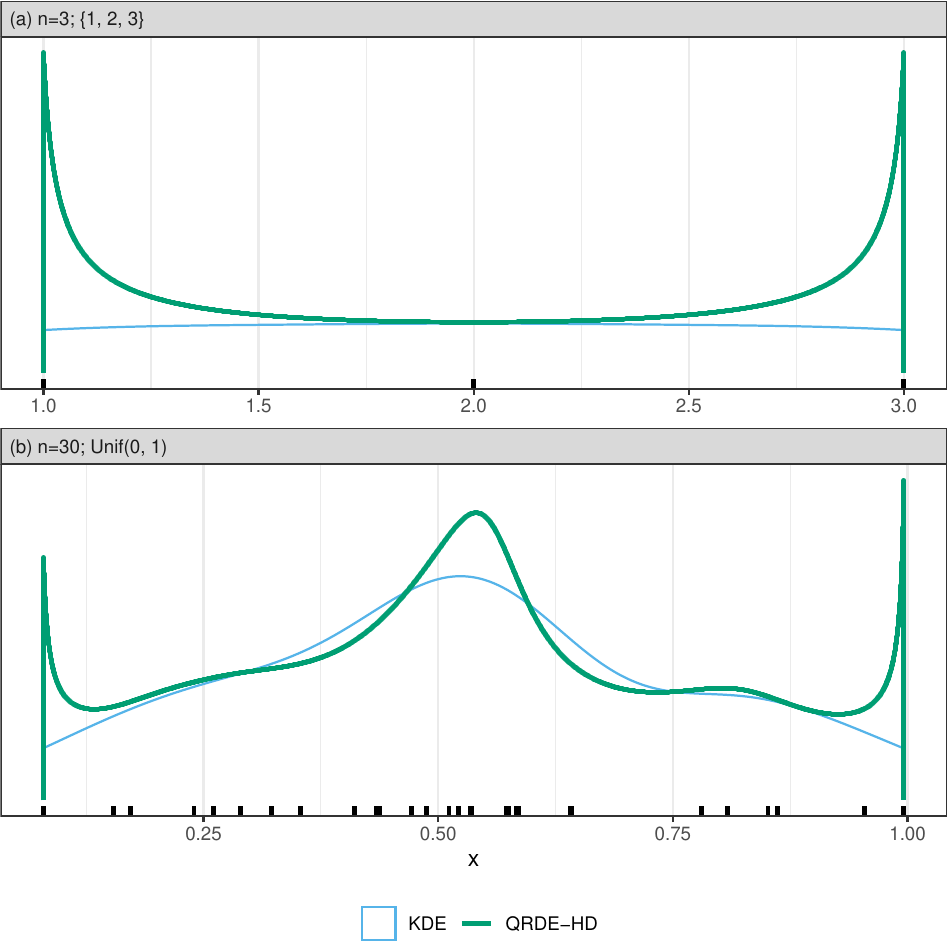} 

}

\caption{QRDE-HD and boundary spikes}\label{fig:tail}
\end{figure}

\end{example}

\clearpage

\begin{example}[QRDE-HD vs. Equal-width Histogram vs. KDE]
\protect\hypertarget{exm:hist-kde}{}\label{exm:hist-kde}

In Figure~\ref{fig:hist-kde}, we compare three density estimators: QRDE-HD, equal-width histogram, and KDE.

In Figure~\ref{fig:hist-kde} (a), we have a bimodal distribution.
KDE produces an over-smoothed estimation and fails to catch local density peaks.
The histogram accurately highlights high-density areas, but it is not smooth.
QRDE-HD provides a trade-off between the histogram and KDE:
it is smoother than the histogram and more accurate than KDE.

In Figure~\ref{fig:hist-kde} (b), we have a heavy-tailed distribution.
The histogram overemphasizes the outliers, and KDE underemphasizes the mode.
Meanwhile, QRDE-HD properly emphasizes the mode and provides a smooth estimation for the tail.
The only visible problem of QRDE-HD is the boundary spikes, which can be silently ignored.

\begin{figure}[H]

{\centering \includegraphics{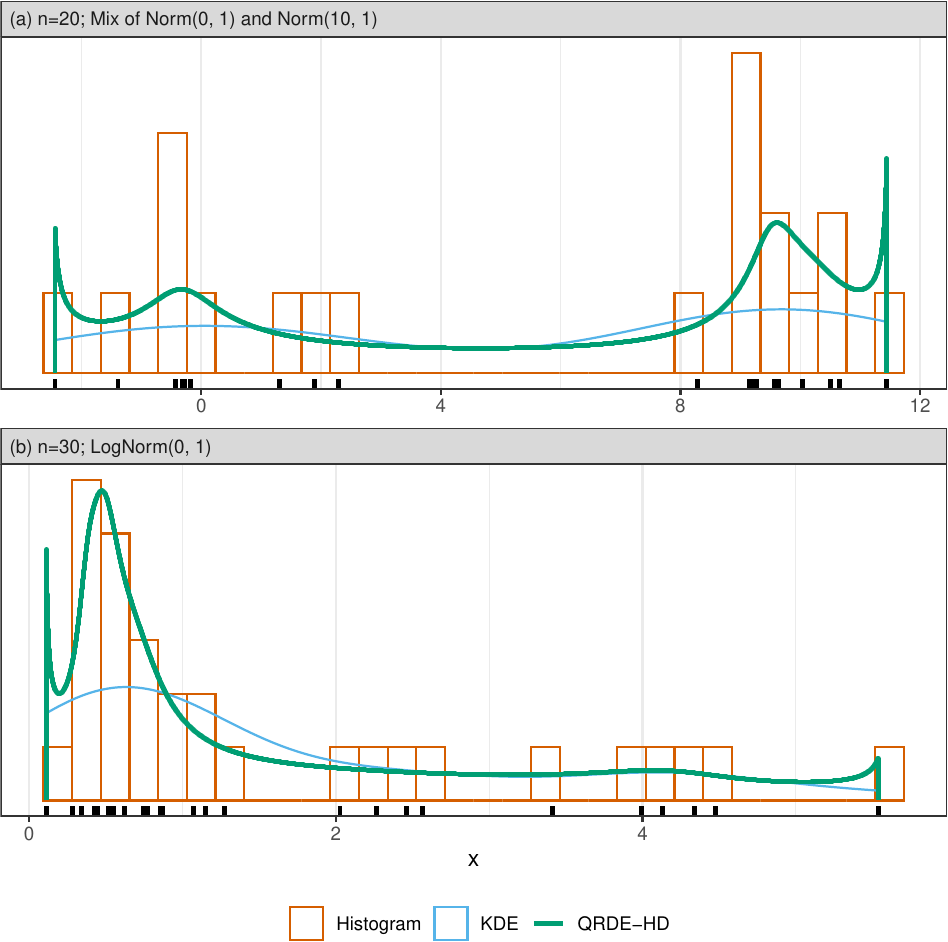} 

}

\caption{QRDE-HD vs. Equal-width Histogram vs. KDE.}\label{fig:hist-kde}
\end{figure}

\end{example}

\clearpage

\begin{example}[QRDE-HD and bimodality]
\protect\hypertarget{exm:bimodality}{}\label{exm:bimodality}

Bimodality is an important feature of a distribution, which is nice to see on a density plot.
In Figure~\ref{fig:bimodality} (a), we have two modes separated by a short gap in the middle of the sample range.
As usual, the default KDE approach fails to properly acknowledge the bimodality effect,
while the histogram and QRDE-HD are capable of doing it.

In Figure~\ref{fig:bimodality} (b), the relative gap between these modes is even smaller.
Since it is smaller than the bin width, the histogram fails to distinguish the modes.
Meanwhile, thanks to the large sample size, QRDE-HD can catch the bimodality effect and show it on the plot.

\begin{figure}[H]

{\centering \includegraphics{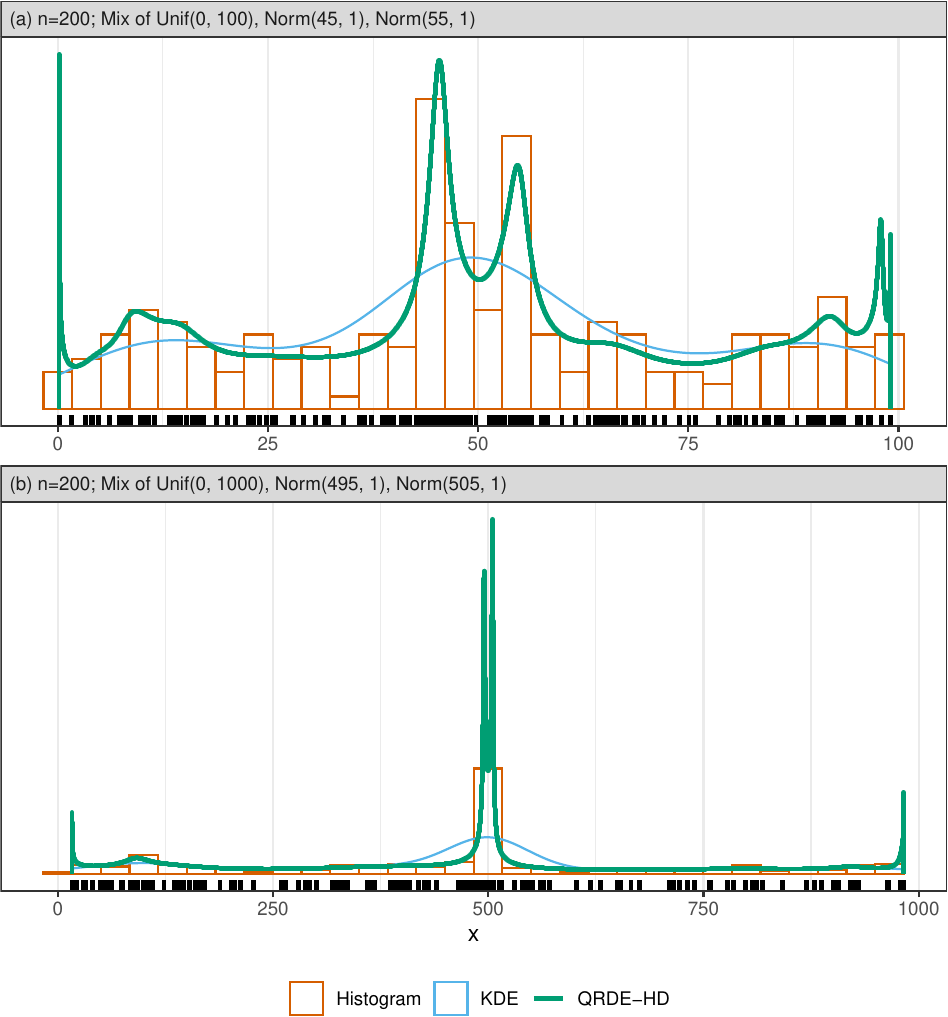} 

}

\caption{QRDE-HD and bimodality.}\label{fig:bimodality}
\end{figure}

\end{example}

\clearpage

\begin{example}[QRDE-HD and multimodality]
\protect\hypertarget{exm:multimodality}{}\label{exm:multimodality}

In Figure~\ref{fig:multimodality}, we have the same three estimators: QRDE-HD, equal-width histogram, and KDE.
However, the number of modes is increased:
we have 10 modes in Figure~\ref{fig:multimodality} (a)
and 30 modes in Figure~\ref{fig:multimodality} (b).
As we can see, only QRDE-HD is capable of detecting all the modes
(assuming the sample is reasonably large to distinguish the modes).

\begin{figure}[H]

{\centering \includegraphics{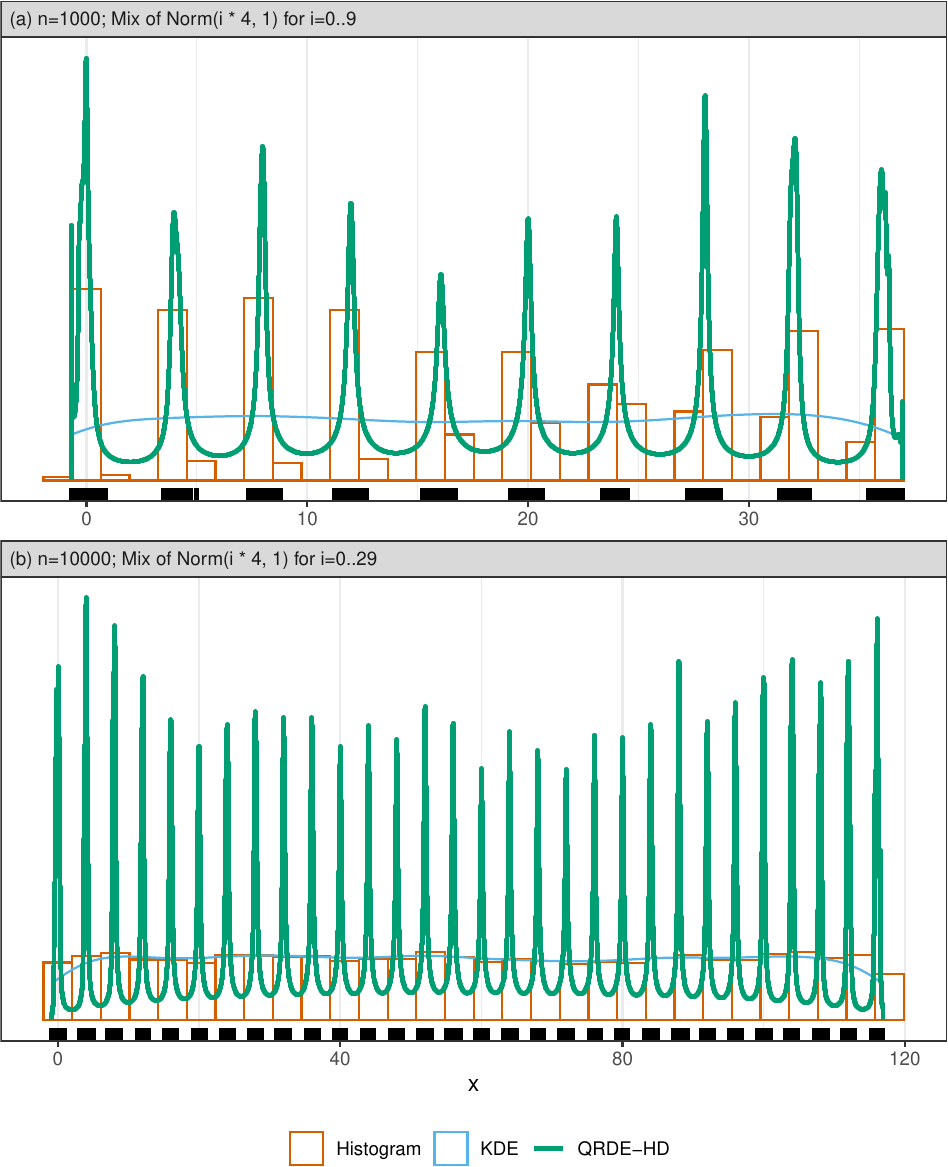} 

}

\caption{QRDE-HD and multimodality.}\label{fig:multimodality}
\end{figure}

\end{example}

\clearpage

\begin{example}[QRDE-HD and overfitting]
\protect\hypertarget{exm:overfitting}{}\label{exm:overfitting}

In Figure~\ref{fig:overfitting}, we see an example of overfitting.
When the sample size is very large, we encounter another disadvantage of QRDE-HD:
it overemphasizes local deviations in the data.
A similar problem is also relevant to the classic KDE,
but QRDE-HD has a stronger tendency to provide overfitted estimations.

\begin{figure}[H]

{\centering \includegraphics{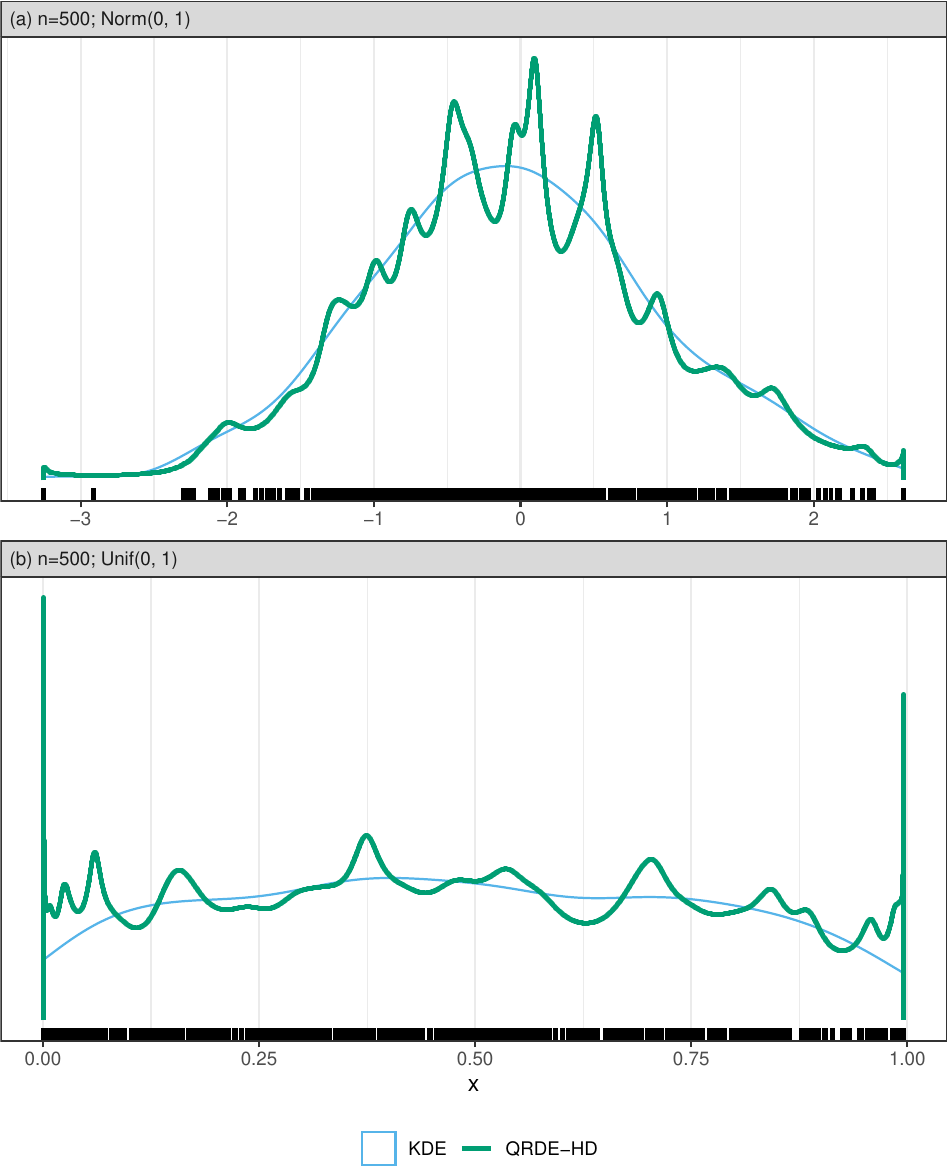} 

}

\caption{QRDE-HD and overfitting.}\label{fig:overfitting}
\end{figure}

\end{example}

\clearpage

\begin{example}[Robustness of QRDE-HD and QRDE-THD]
\protect\hypertarget{exm:robustness}{}\label{exm:robustness}

The Harrell-Davis quantile estimator is not robust and may be affected by outliers.
While practical robustness is good in most cases,
some extreme outliers may significantly distort the whole estimation.
In Figure~\ref{fig:robustness}, we consider two samples:
(a) a sample from \(\mathcal{U}(0, 1)\) of size \(50\) with an added \(10^3\);
(b) the same sample with the outlier replaced by \(10^9\).
In the range \(x \in [0.82;0.95]\), we can observe the dynamic of change in the QRDE-HD values.

A robust alternative to the classic \(\QHD\) is
its trimmed modification \(\QTHD\) (see \autocite{akinshin2022thdqe})
which noticeably increases the robustness, sacrificing a small portion of the efficiency.
The proposed rule of thumb for the size of the trimmed interval is \(\sqrt{n}\),
which should be used \emph{only} in the trial experiments or in the absence of any prior knowledge of the domain area.
It is recommended that a justified choice be made based on research requirements for robustness and efficiency.
We can build a QRDE based on \(\QTHD\) (denoted by QRDE-THD) in the same way as we build QRDE-HD.
The benefit of QRDE-THD is better robustness and higher resistance to outliers.
Unfortunately, the downside is the lack of smoothness.
In Figure~\ref{fig:robustness}, QRDE-THD is presented in contrast to QRDE-HD.
As we can see, its plot is not smooth, but it more accurately describes the density.
It is recommended to use QRDE-THD when the magnitude of the deviations is several degrees larger than
the dispersion of the sample primary range.

\begin{figure}[H]

{\centering \includegraphics{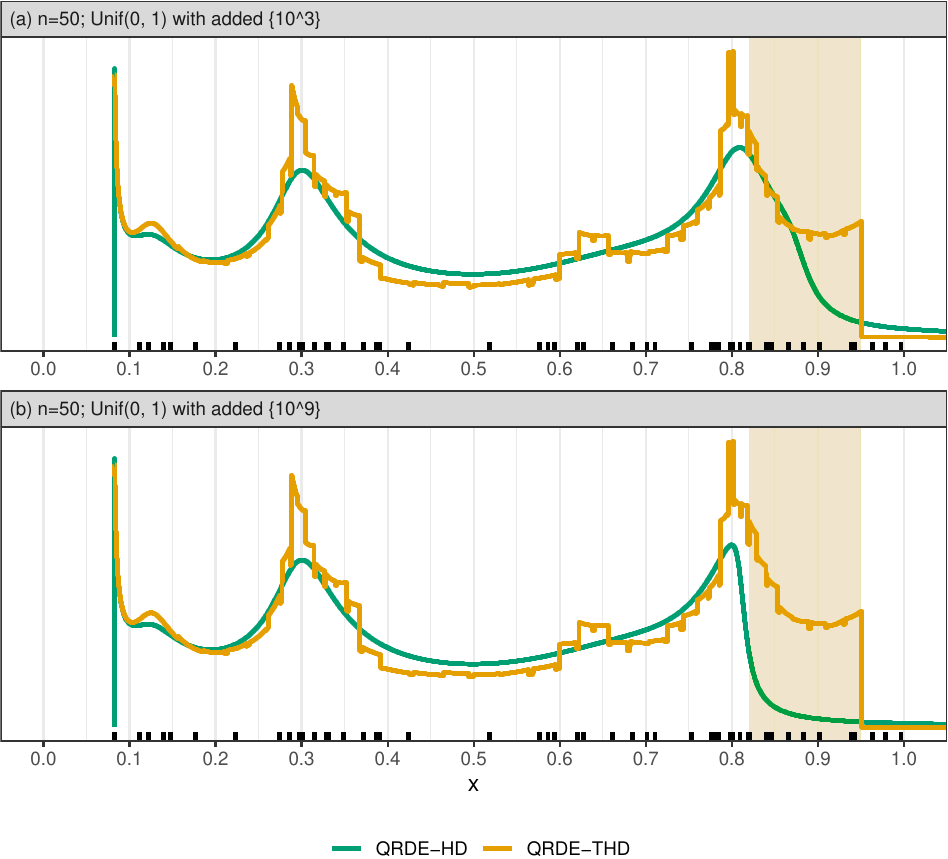} 

}

\caption{QRDE-HD and QRDE-THD in the presence of outliers.}\label{fig:robustness}
\end{figure}

\end{example}

\clearpage

\section{Jittering}\label{sec:jit}

Density estimation is designed to describe continuous distributions.
With such a mathematical model, we have an implicit assumption that the probability of observing tied values is zero.
Meanwhile, the resolution of the measurement tools is always limited by a finite value,
which means that the actually collected observations are essentially discrete.
However, when the resolution is much smaller than typical gaps between observations,
treating all the distributions as discrete is impractical.
The use of continuous models usually provides a decent approximation and more powerful analysis approaches.

When we start reducing the resolution, the discretization effect may appear in the form of occasional tied values.
In many real-life applications, this effect is not strong enough to switch to a discrete model,
but it prevents us from relying on the ``no tied values possible'' assumption.
Therefore, if we want to build a practically reliable density estimation,
we have to handle the tied values properly.
Let us review Example~\ref{exm:ties} and Example~\ref{exm:discretization} to better understand the problem.

\begin{example}[Density estimation and tied values]
\protect\hypertarget{exm:ties}{}\label{exm:ties}Let us consider the following sample:

\[
\x = (1,\, 1.9,\, 2,\, 2.1,\, 3).
\]

If we estimate the traditional quartiles using \(\QHFS\), we will get a convenient result:

\[
\QHFS(\x, 0)    = 1,\quad
\QHFS(\x, 0.25) = 1.9,\quad
\QHFS(\x, 0.5)  = 2,\quad
\QHFS(\x, 0.75) = 2.1,\quad
\QHFS(\x, 1)    = 3.
\]

We consider a density estimation using the suggested pseudo-histogram approximation approach with \(k=4\) (\(\xi=0.25\)).
Instead of building the whole histogram, we focus on the second bin.
Its height can be easily calculated:

\[
h_2 =
\frac{\xi}{\QHFS(\x, 2\xi)-\QHFS(\x, 1\xi)} =
\frac{0.25}{\QHFS(\x, 0.5)-\QHFS(\x, 0.25)} =
\frac{0.25}{2-1.9} = 2.5.
\]

Now, let us introduce a rounded version of \(\x\):

\[
\x^{\circ} = \operatorname{Round}(\x) = (1,\, 2,\, 2,\, 2,\, 3).
\]

The quantile values will be correspondingly rounded:

\[
\QHFS(\x^{\circ}, 0)    = 1,\quad
\QHFS(\x^{\circ}, 0.25) = 2,\quad
\QHFS(\x^{\circ}, 0.5)  = 2,\quad
\QHFS(\x^{\circ}, 0.75) = 2,\quad
\QHFS(\x^{\circ}, 1)    = 3.
\]

Unfortunately, when we start building the pseudo-histogram, we will get a problem:

\[
h^{\circ}_2 =
\frac{\xi}{\QHFS(\x^{\circ}, 2\xi)-\QHFS(\x^{\circ}, 1\xi)} =
\frac{0.25}{\QHFS(\x^{\circ}, 0.5)-\QHFS(\x^{\circ}, 0.25)} =
\frac{0.25}{0}.
\]

In the continuous world, we cannot define the distribution density at the point where two different quantiles are equal.
Switching from \(\QHFS\) to \(\QHD\) improves the situation
since all the \(\QHD\) estimations are different, and we avoid division by zero.
However, if we have multiple tied values, \(\QHD\) estimations may become too close to each other,
which will also lead to unreasonably large spikes in the density estimation.
\end{example}

\clearpage

\begin{example}[QRDE-HD and discretization]
\protect\hypertarget{exm:discretization}{}\label{exm:discretization}

Let us consider a sample of size \(2000\) from the standard normal distribution.
We round all the measurements to the first decimal digit
and build KDE and QRDE-HD which are presented in Figure~\ref{fig:discretization} (a).
While KDE provides a smooth estimation of the normal distribution,
QRDE-HD produces spikes at the tied values.
The correctness of such a representation is a philosophical question.
Indeed, when should we stop considering data as a smooth continuous model
and start treating it as a mixture of multiple Dirac delta functions?
There is no universal answer: everything depends on the problem and the research goals.

Meanwhile, the discretization effect is relevant not only to QRDE-HD but also to other density estimators.
In Figure~\ref{fig:discretization} (b),
we divide the original sample by \(5\), round it to the first decimal digit, and build a KDE.
As we can see, with a higher number of tied values, KDE also experiences the same phenomenon.
The only difference between QRDE-HD and KDE is the tipping point at which the discretization effect becomes noticeable.
QRDE-HD has higher sensitivity to ties, which is beneficial for multimodality detection
but can be considered a disadvantage in the case of discretization.

\begin{figure}[H]

{\centering \includegraphics{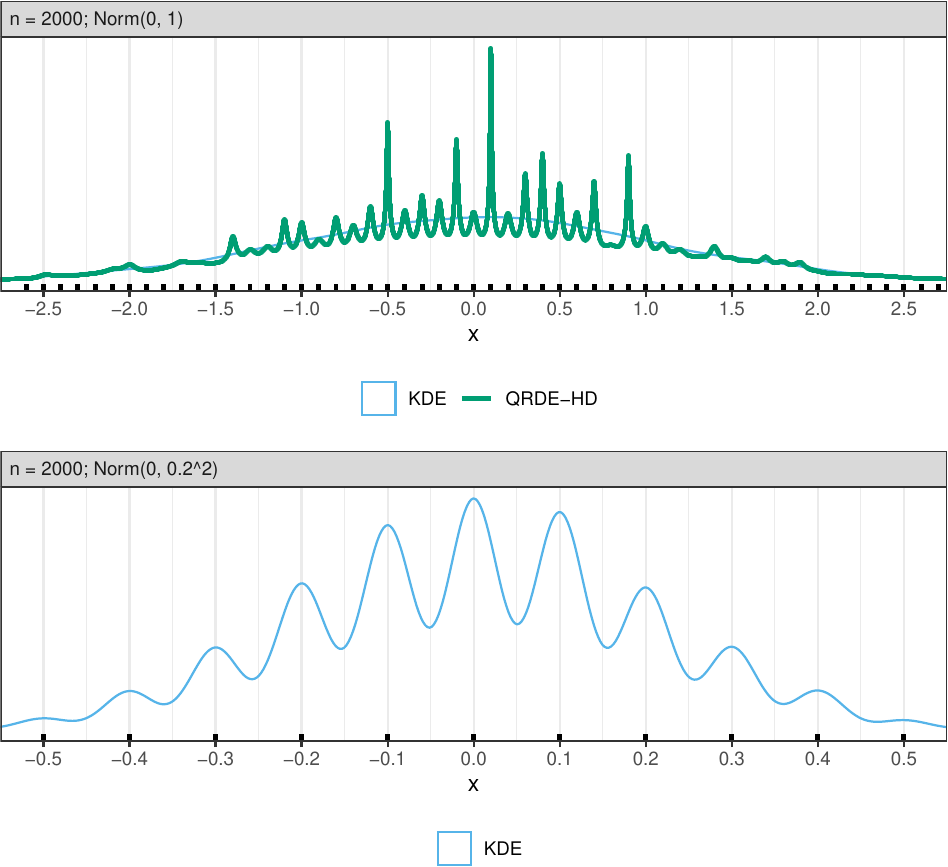} 

}

\caption{KDE and discretization.}\label{fig:discretization}
\end{figure}

\end{example}

\clearpage

While Example~\ref{exm:ties} and Example~\ref{exm:discretization}
illustrate an essential limitation of the density estimation approach,
they also highlight an important disadvantage of the straightforward QRDE-HD approach.
Imagine that we have a sample without tied values and the corresponding QRDE-HD.
Next, we decide that the precision of obtained measurements is too high: the last digits of each value are just noise.
Therefore, we round the measurements, which leads to tied values.
If we are correct and the dropped digits are non-meaningful noise,
the rounding procedure does not lead to information loss.
Essentially, we still have the same sample.
However, the emergence of tied values may prevent us from building reasonably-looking QRDE-HD.
That reduces the practical applicability of the suggested approach.

In order to make the density estimation resilient to tied values,
we suggest adding noise to the data points in order to spread identical measurements apart.
This technique (which is known as jittering, see \autocite{nagler2018a}) may feel controversial.
While statisticians spend a lot of effort trying to eliminate the noise,
jittering suggests distorting the data on purpose.
One may have concerns related to the possible efficiency loss or bias introduction.
In \autocite{nagler2018b}, a proper justification is provided:
it is claimed that jittering does not have a negative impact on the estimation accuracy.
According to this work, jittered estimators are not expected to have practically significant deviations
from their non-jittered prototypes.
While this statement tends to be true in practice, it is worth discussing the details.
There are multiple ways to implement jittering.
The proper research focus should be on the specific jittering algorithm,
not on the concept of jittering in general.
In order to choose one, we should define the desired properties first.
We suggest the following list of requirements:

\medskip

\begin{itemize}
\tightlist
\item
  \textbf{Do not extend the sample range}\\
  The assumption of zero density outside the sample range may be required for further statistical procedure.
  In order to make jittering applicable for a wider range of use cases,
  it would be beneficial if we preserve the sample range.
  For example, if the actual data elements are non-negative by nature and there are tied zero values,
  a random noise will introduce negative elements,
  which may be out of the support area for some equations
  (imagine that a lower percentile value is used inside a square root).
  Why would we introduce negative elements in such cases
  if we have an option to change the noise shape and preserve the sample range?
\item
  \textbf{Use deterministic jittering}\\
  Deterministic behavior is also preferable over non-deterministic if all the other conditions are the same.
  Any kind of randomization may become a source of flakiness and reduce the reproducibility of analysis.
  Unlike other statistical methods like bootstrap or Monte-Carlo
  that intentionally exploit the properties of random distributions,
  we do not have such a need.
  The only goal here is to get rid of the tied values.
  Since the noise is supposed to be negligible, we are free to choose any kind of noise shape.
  The exact noise pattern can be designed in advance to ensure
  identical QRDE-HD charts in case of multiple usages of the same data.
\item
  \textbf{Apply jittering only to the tied values}\\
  Jittering, in general, does not mean that we should add noise to \emph{all} sample elements.
  It is reasonable to limit the scope of jittering only to the tied values
  and avoid unnecessary distortion of unique sample elements.
\item
  \textbf{The actual resolution from the domain area should be acknowledged}\\
  In order to implement jittering, we should pick up the noise magnitude.
  We recommend using half of the measure resolution as the maximum deviation from the original value.
  An approach of estimating the distortion range adaptively based on the observed data is not always applicable.
  Let us consider a sample of four elements: \(\mathbf{x} = (1, 1, 2, 2)\).
  How should we present it?
  If we are talking about millimeters and \(\mathbf{x}\) was obtained using a ruler with a resolution of 1mm,
  the noise magnitude of 0.5mm tends to lead to a uniform density or even to a slightly bell-shaped density,
  which better describes our probable expectations of a high density of around 1.5mm.
  If we are talking about kilometers and \(\mathbf{x}\) contains physical distances
  measured with a max error of 20 meters,
  the noise magnitude of 10 meters will lead to a bimodal density with modes at 1km and 2km.
  Deriving the noise range from the domain area rather than from the sample
  (e.g., a minimum non-zero distance between sample elements)
  helps to avoid confusion in corner cases like the one mentioned above.
\end{itemize}

\clearpage

Such scrutiny may be perceived as unnecessarily elevating minor concerns.
However, we believe that it is important to pay due attention to detail.
If we can satisfy these requirements for free without any drawbacks and make the implementation applicable
to a wider range of corner cases that have a chance to appear in real data sets,
we do not observe reasons to avoid this opportunity.
We also provide a simple ready-to-use implementation in Appendix~\ref{sec:refimpl},
so that the reader can run it on their data and check if it performs well.

Let us propose a possible noise pattern that satisfies the above requirements.
Since we want to prevent noticeable gaps between jittered measurements, it feels reasonable to try the uniform noise.
Let \(\x_{(i:j)}\) be a range of tied order statistics of width \(k=j-i+1\).
We want to define a noise vector \(\xi_{i:j}\) to obtain the jittered sample \(\acute{\x}\) which is defined by

\[
\acute{x}_{(i:j)} = \x_{(i:j)} + \xi_{i:j}.
\]

Let \(u\) be a vector of indexes \((1,2,\ldots,n)\) linearly mapped to the range \([0;1]\):

\[
u_i = (i - 1) / (k - 1),\quad \textrm{for}\quad i = 1, 2, \ldots, k.
\]

We suggest spreading the \(\x_{(i:j)}\) values apart using the following rules:

\medskip

\begin{itemize}
\tightlist
\item
  If (\(i > 1\) and \(j < n\)), we use \(\xi_{i:j} = s \cdot (u_{i:j} - 0.5)\).\\
  In order to support the zero dispersion case, we also use this rule if (\(i = 1\) and \(j = n\)).
\item
  If (\(i = 1\) and \(j < n\)), we use \(\xi_{i:j} = s \cdot (u_{i:j} / 2)\).\\
  Therefore, at the tied values at the minimum value, we extend the range to the right.
\item
  If (\(i > 1\) and \(j = n\)), we use \(\xi_{i:j} = s \cdot (u_{i:j} - 1) / 2\).\\
  Therefore, at the tied values at the maximum value, we extend the range to the left.
\end{itemize}

\medskip

The suggested approach preserves the sample range, provides a small bias, and returns consistent non-randomized values.
The only case of range extension is a sample of zero-width range,
which does not allow for the building of a reasonable density plot preserving equal minimum and maximum values.
The knowledge of the resolution value \(s\) helps to guarantee the absence of reordering.

The uniform approach works quite well and allows pseudo-restoring of the original noise
which is required to protect QRDE-HD from the discretization effect.

Let us consider one more example to show how jittering can restore original density estimation
after rounding the sample values.

\clearpage

\begin{example}[Jittering in action]
\protect\hypertarget{exm:jittering}{}\label{exm:jittering}

In Figure~\ref{fig:jittering} (a),
we can see a QRDE-HD for a sample of 2000 elements from the standard normal distribution.
In Figure~\ref{fig:jittering} (b),
we can see a QRDE-HD for a rounded version of the same sample up to the first decimal digit.
In Figure~\ref{fig:jittering} (c),
we can see a QRDE-HD for a jittered version of the rounded sample.
In Figure~\ref{fig:jittering} (d),
we can see a comparison of the QRDE-HD for the original and jittered sample.
It is easy to see that jittering restored the initial density with minor deviations
and helped us overcome the problem of the tied values.

\begin{figure}[H]

{\centering \includegraphics{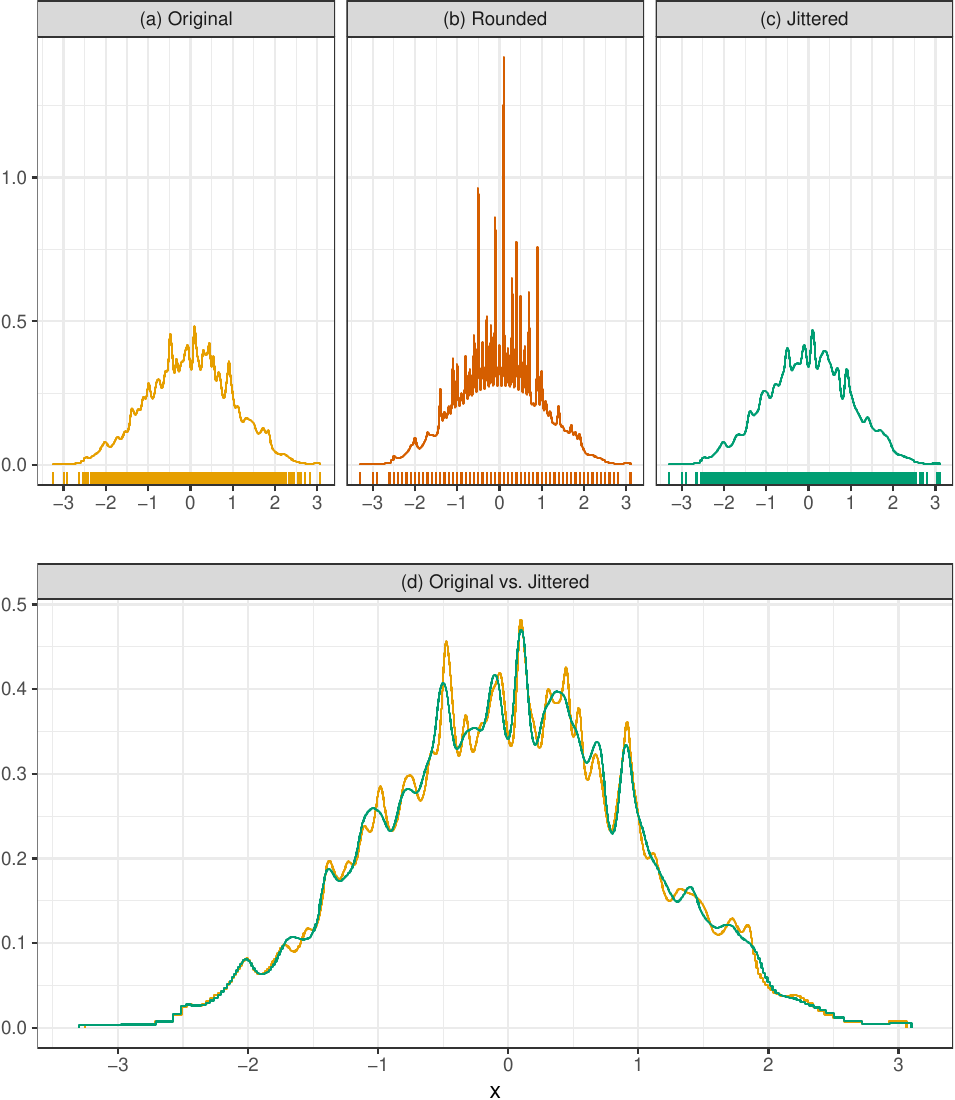} 

}

\caption{Density plots for a sample from Norm(0, 1), its rounded and jittered versions.}\label{fig:jittering}
\end{figure}

\end{example}

\clearpage

\section{Summary}\label{sec:summary}

In this paper, we propose a new density estimator QRDE-HD which is consistent with the Harrell-Davis quantile estimator.
This approach has the following advantages:

\begin{itemize}
  \item[\textcolor{good}{\textbullet}] QRDE-HD can be simultaneously used with a complementary quantile estimator.
  \item[\textcolor{good}{\textbullet}] QRDE-HD accurately emphasizes multimodality.
  \item[\textcolor{good}{\textbullet}] QRDE-HD is always smooth.
  \item[\textcolor{good}{\textbullet}] QRDE-HD never extends the sample range.
  \item[\textcolor{good}{\textbullet}] QRDE-HD automatically adjusts to the data
                                         without a need to be tuned via a bandwidth parameter.
  \item[\textcolor{good}{\textbullet}] QRDE-HD provides a good trade-off between
                                         the non-smooth histogram and over-smoothed KDE.
\end{itemize}

\medskip

However, QRDE-HD has its area of applicability.
Since it does not fit all the problems, it is important to highlight its disadvantages:

\begin{itemize}
  \item[\textcolor{bad}{\textbullet}] QRDE-HD inaccurately approximates tails and produce boundary spikes.
  \item[\textcolor{bad}{\textbullet}] QRDE-HD suffers from underfitting on
                                        small samples and overfitting on large samples.
  \item[\textcolor{bad}{\textbullet}] QRDE-HD is not robust and can be affected by extreme outliers
                                        (can be resolved using QRDE-THD).
  \item[\textcolor{bad}{\textbullet}] QRDE-HD may produce huge spikes on tied values (can be resolved using jittering).
\end{itemize}

\medskip

While we do not recommend thoughtlessly applying QRDE to all possible problems,
we believe that it is a valuable addition to the existing set of density estimators,
which may be beneficial when the classic methods do not perform well enough.

\section*{Disclosure statement}\label{disclosure-statement}
\addcontentsline{toc}{section}{Disclosure statement}

The author declares no conflict of interest.

\section*{Data and source code availability}\label{data-and-source-code-availability}
\addcontentsline{toc}{section}{Data and source code availability}

The source code of this paper and all the examples are available on GitHub:\\
\url{https://github.com/AndreyAkinshin/paper-qrdehd}.

\section*{Acknowledgments}\label{acknowledgments}
\addcontentsline{toc}{section}{Acknowledgments}

The author thanks Ivan Pashchenko for valuable discussions.

\clearpage

\appendix

\section{Reference implementation}\label{sec:refimpl}

Below are R\footnote{\url{https://www.r-project.org/}} implementations of a ggplot2\footnote{\url{https://ggplot2.tidyverse.org/}}-based QRDE-HD plotting function and a jittering function.

\begin{Shaded}
\begin{Highlighting}[]
\FunctionTok{library}\NormalTok{(ggplot2) }\CommentTok{\# Plotting}
\FunctionTok{library}\NormalTok{(Hmisc)   }\CommentTok{\# Harrell{-}Davis quantile estimator}

\NormalTok{StatQDensity }\OtherTok{\textless{}{-}} \FunctionTok{ggproto}\NormalTok{(}\StringTok{"StatQDensity"}\NormalTok{, Stat,}
  \AttributeTok{compute\_group =} \ControlFlowTok{function}\NormalTok{(data, scales, bincount, Q) \{}
    \CommentTok{\# Transforming the input data$x to QRDE{-}HD}
\NormalTok{    p }\OtherTok{\textless{}{-}} \FunctionTok{seq}\NormalTok{(}\DecValTok{0}\NormalTok{, }\DecValTok{1}\NormalTok{, }\AttributeTok{length.out =}\NormalTok{ bincount }\SpecialCharTok{+} \DecValTok{1}\NormalTok{)}
\NormalTok{    q }\OtherTok{\textless{}{-}} \FunctionTok{Q}\NormalTok{(data}\SpecialCharTok{$}\NormalTok{x, p)}
\NormalTok{    h }\OtherTok{\textless{}{-}} \FunctionTok{pmax}\NormalTok{(}\DecValTok{1} \SpecialCharTok{/}\NormalTok{ bincount }\SpecialCharTok{/}\NormalTok{ (}\FunctionTok{tail}\NormalTok{(q, }\SpecialCharTok{{-}}\DecValTok{1}\NormalTok{) }\SpecialCharTok{{-}} \FunctionTok{head}\NormalTok{(q, }\SpecialCharTok{{-}}\DecValTok{1}\NormalTok{)), }\DecValTok{0}\NormalTok{)}
\NormalTok{    den\_x }\OtherTok{\textless{}{-}} \FunctionTok{rep}\NormalTok{(q, }\AttributeTok{each =} \DecValTok{2}\NormalTok{)}
\NormalTok{    den\_y }\OtherTok{\textless{}{-}} \FunctionTok{c}\NormalTok{(}\DecValTok{0}\NormalTok{, }\FunctionTok{rep}\NormalTok{(h, }\AttributeTok{each =} \DecValTok{2}\NormalTok{), }\DecValTok{0}\NormalTok{)}
    \FunctionTok{data.frame}\NormalTok{(}\AttributeTok{x =}\NormalTok{ den\_x, }\AttributeTok{y =}\NormalTok{ den\_y)}
\NormalTok{  \}}
\NormalTok{)}

\CommentTok{\#\textquotesingle{} @param bincount the number of bins in the pseudo{-}histogram}
\CommentTok{\#\textquotesingle{} @param Q the target quantile estimator (default: Harrell{-}Davis)}
\NormalTok{geom\_qrdensity }\OtherTok{\textless{}{-}} \ControlFlowTok{function}\NormalTok{(}\AttributeTok{mapping =} \ConstantTok{NULL}\NormalTok{, }\AttributeTok{data =} \ConstantTok{NULL}\NormalTok{,}
                           \AttributeTok{stat =} \StringTok{"qdensity"}\NormalTok{, }\AttributeTok{position =} \StringTok{"identity"}\NormalTok{,}
                           \AttributeTok{bincount =} \DecValTok{1000}\NormalTok{, }\AttributeTok{Q =}\NormalTok{ hdquantile, ...) \{}
  \FunctionTok{layer}\NormalTok{(}
    \AttributeTok{stat =}\NormalTok{ StatQDensity,}
    \AttributeTok{data =}\NormalTok{ data,}
    \AttributeTok{mapping =}\NormalTok{ mapping,}
    \AttributeTok{geom =}\NormalTok{ GeomLine,}
    \AttributeTok{position =}\NormalTok{ position,}
    \AttributeTok{params =} \FunctionTok{list}\NormalTok{(}\AttributeTok{bincount =}\NormalTok{ bincount, }\AttributeTok{Q =}\NormalTok{ Q, ...),}
\NormalTok{  )}
\NormalTok{\}}

\CommentTok{\# Demo}
\FunctionTok{set.seed}\NormalTok{(}\DecValTok{42}\NormalTok{)}
\NormalTok{x }\OtherTok{\textless{}{-}} \FunctionTok{numeric}\NormalTok{(}\DecValTok{200}\NormalTok{)}
\NormalTok{mix }\OtherTok{\textless{}{-}} \FunctionTok{sample}\NormalTok{(}\FunctionTok{c}\NormalTok{(}\FunctionTok{rep}\NormalTok{(}\DecValTok{0}\NormalTok{, }\DecValTok{8}\NormalTok{), }\DecValTok{1}\NormalTok{, }\DecValTok{2}\NormalTok{), }\DecValTok{200}\NormalTok{, }\ConstantTok{TRUE}\NormalTok{)}
\NormalTok{x[mix }\SpecialCharTok{==} \DecValTok{0}\NormalTok{] }\OtherTok{\textless{}{-}} \FunctionTok{runif}\NormalTok{(}\FunctionTok{sum}\NormalTok{(mix }\SpecialCharTok{==} \DecValTok{0}\NormalTok{), }\DecValTok{0}\NormalTok{, }\DecValTok{100}\NormalTok{)}
\NormalTok{x[mix }\SpecialCharTok{==} \DecValTok{1}\NormalTok{] }\OtherTok{\textless{}{-}} \FunctionTok{rnorm}\NormalTok{(}\FunctionTok{sum}\NormalTok{(mix }\SpecialCharTok{==} \DecValTok{1}\NormalTok{), }\DecValTok{45}\NormalTok{)}
\NormalTok{x[mix }\SpecialCharTok{==} \DecValTok{2}\NormalTok{] }\OtherTok{\textless{}{-}} \FunctionTok{rnorm}\NormalTok{(}\FunctionTok{sum}\NormalTok{(mix }\SpecialCharTok{==} \DecValTok{2}\NormalTok{), }\DecValTok{55}\NormalTok{)}
\FunctionTok{ggplot}\NormalTok{(}\FunctionTok{data.frame}\NormalTok{(x), }\FunctionTok{aes}\NormalTok{(x)) }\SpecialCharTok{+}
  \FunctionTok{geom\_histogram}\NormalTok{(}\FunctionTok{aes}\NormalTok{(}\AttributeTok{y =} \FunctionTok{after\_stat}\NormalTok{(density)), }\AttributeTok{bins =} \DecValTok{30}\NormalTok{,}
                 \AttributeTok{fill =} \StringTok{"transparent"}\NormalTok{, }\AttributeTok{col =} \StringTok{"black"}\NormalTok{) }\SpecialCharTok{+}
  \FunctionTok{geom\_density}\NormalTok{(}\AttributeTok{col =} \StringTok{"red"}\NormalTok{) }\SpecialCharTok{+}
  \FunctionTok{geom\_qrdensity}\NormalTok{(}\AttributeTok{col =} \StringTok{"\#00AA00"}\NormalTok{, }\AttributeTok{linewidth =} \FloatTok{1.2}\NormalTok{) }\SpecialCharTok{+}
  \FunctionTok{geom\_rug}\NormalTok{(}\AttributeTok{sides =} \StringTok{"b"}\NormalTok{)}
\end{Highlighting}
\end{Shaded}

\clearpage

\begin{Shaded}
\begin{Highlighting}[]
\CommentTok{\#\textquotesingle{} @param x sample}
\CommentTok{\#\textquotesingle{} @param s resolution of the measurements}
\NormalTok{jitter }\OtherTok{\textless{}{-}} \ControlFlowTok{function}\NormalTok{(x, s) \{}
\NormalTok{  x }\OtherTok{\textless{}{-}} \FunctionTok{sort}\NormalTok{(x)}
\NormalTok{  n }\OtherTok{\textless{}{-}} \FunctionTok{length}\NormalTok{(x)}
  \CommentTok{\# Searching for intervals [i;j] of tied values}
\NormalTok{  i }\OtherTok{\textless{}{-}} \DecValTok{1}
  \ControlFlowTok{while}\NormalTok{ (i }\SpecialCharTok{\textless{}=}\NormalTok{ n) \{}
\NormalTok{    j }\OtherTok{\textless{}{-}}\NormalTok{ i}
    \ControlFlowTok{while}\NormalTok{ (j }\SpecialCharTok{\textless{}}\NormalTok{ n }\SpecialCharTok{\&\&}\NormalTok{ x[j }\SpecialCharTok{+} \DecValTok{1}\NormalTok{] }\SpecialCharTok{{-}}\NormalTok{ x[i] }\SpecialCharTok{\textless{}}\NormalTok{ s }\SpecialCharTok{/} \DecValTok{2}\NormalTok{) \{}
\NormalTok{      j }\OtherTok{\textless{}{-}}\NormalTok{ j }\SpecialCharTok{+} \DecValTok{1}
\NormalTok{    \}}
    \ControlFlowTok{if}\NormalTok{ (i }\SpecialCharTok{\textless{}}\NormalTok{ j }\SpecialCharTok{\&\&}\NormalTok{ j }\SpecialCharTok{{-}}\NormalTok{ i }\SpecialCharTok{+} \DecValTok{1} \SpecialCharTok{\textless{}}\NormalTok{ n) \{}
\NormalTok{      k }\OtherTok{\textless{}{-}}\NormalTok{ j }\SpecialCharTok{{-}}\NormalTok{ i }\SpecialCharTok{+} \DecValTok{1}
\NormalTok{      u }\OtherTok{\textless{}{-}} \DecValTok{0}\SpecialCharTok{:}\NormalTok{(k }\SpecialCharTok{{-}} \DecValTok{1}\NormalTok{) }\SpecialCharTok{/}\NormalTok{ (k }\SpecialCharTok{{-}} \DecValTok{1}\NormalTok{)}
\NormalTok{      xi }\OtherTok{\textless{}{-}}\NormalTok{ u }\SpecialCharTok{{-}} \FloatTok{0.5}
      \ControlFlowTok{if}\NormalTok{ (i }\SpecialCharTok{==} \DecValTok{1}\NormalTok{)}
\NormalTok{        xi }\OtherTok{\textless{}{-}}\NormalTok{ u }\SpecialCharTok{/} \DecValTok{2}
      \ControlFlowTok{if}\NormalTok{ (j }\SpecialCharTok{==}\NormalTok{ n)}
\NormalTok{        xi }\OtherTok{\textless{}{-}}\NormalTok{ (u }\SpecialCharTok{{-}} \DecValTok{1}\NormalTok{) }\SpecialCharTok{/} \DecValTok{2}
      \ControlFlowTok{if}\NormalTok{ (i }\SpecialCharTok{==} \DecValTok{1} \SpecialCharTok{\&\&}\NormalTok{ j }\SpecialCharTok{==}\NormalTok{ n)}
\NormalTok{        xi }\OtherTok{\textless{}{-}}\NormalTok{ u }\SpecialCharTok{{-}} \FloatTok{0.5}
\NormalTok{      x[i}\SpecialCharTok{:}\NormalTok{j] }\OtherTok{\textless{}{-}}\NormalTok{ x[i}\SpecialCharTok{:}\NormalTok{j] }\SpecialCharTok{+}\NormalTok{ xi }\SpecialCharTok{*}\NormalTok{ s}
    
\NormalTok{    \}}
\NormalTok{    i }\OtherTok{\textless{}{-}}\NormalTok{ j }\SpecialCharTok{+} \DecValTok{1}
\NormalTok{  \}}
  \FunctionTok{return}\NormalTok{(x)}
\NormalTok{\}}

\CommentTok{\# Demo}
\FunctionTok{set.seed}\NormalTok{(}\DecValTok{1729}\NormalTok{)}
\NormalTok{x }\OtherTok{\textless{}{-}} \FunctionTok{rnorm}\NormalTok{(}\DecValTok{2000}\NormalTok{)}
\NormalTok{xr }\OtherTok{\textless{}{-}} \FunctionTok{round}\NormalTok{(x, }\DecValTok{1}\NormalTok{)}
\NormalTok{xj }\OtherTok{\textless{}{-}} \FunctionTok{jitter}\NormalTok{(xr, }\FloatTok{0.1}\NormalTok{)}
\NormalTok{df }\OtherTok{\textless{}{-}} \FunctionTok{rbind}\NormalTok{(}
  \FunctionTok{data.frame}\NormalTok{(}\AttributeTok{type =} \StringTok{"(a) Original"}\NormalTok{, }\AttributeTok{x =}\NormalTok{ x),}
  \FunctionTok{data.frame}\NormalTok{(}\AttributeTok{type =} \StringTok{"(b) Rounded"}\NormalTok{, }\AttributeTok{x =}\NormalTok{ xr),}
  \FunctionTok{data.frame}\NormalTok{(}\AttributeTok{type =} \StringTok{"(c) Jittered"}\NormalTok{, }\AttributeTok{x =}\NormalTok{ xj)}
\NormalTok{)}
\FunctionTok{ggplot}\NormalTok{(df, }\FunctionTok{aes}\NormalTok{(x)) }\SpecialCharTok{+}
  \FunctionTok{facet\_wrap}\NormalTok{(}\FunctionTok{vars}\NormalTok{(type), }\AttributeTok{nrow =} \DecValTok{1}\NormalTok{) }\SpecialCharTok{+}
  \FunctionTok{geom\_qrdensity}\NormalTok{() }\SpecialCharTok{+}
  \FunctionTok{geom\_rug}\NormalTok{(}\AttributeTok{sides =} \StringTok{"b"}\NormalTok{)}
\end{Highlighting}
\end{Shaded}

\begin{Shaded}
\begin{Highlighting}[]
\FunctionTok{ggplot}\NormalTok{(df[df}\SpecialCharTok{$}\NormalTok{type }\SpecialCharTok{!=} \StringTok{"(b) Rounded"}\NormalTok{,], }\FunctionTok{aes}\NormalTok{(x, }\AttributeTok{col =}\NormalTok{ type)) }\SpecialCharTok{+}
  \FunctionTok{geom\_qrdensity}\NormalTok{()}
\end{Highlighting}
\end{Shaded}

\newpage

\printbibliography

\end{document}